\input{aipcheck}

\documentclass[
    ,final            
  ]
  {aipproc}

\layoutstyle{6x9}

\begin{document}

\title{Detailed abundances in stars belonging to ultra-faint dwarf spheroidal galaxies}

\classification{ 98.52.Wz - 98.62.Bj - 97.10.Tk - 97.20.Tr}
\keywords      {Dwarf galaxies - Stars - Abundances }

\author{P. Fran\c{c}ois}{
  address={2GEPI, Observatoire de Paris, CNRS, Univ. Paris Diderot, 61 Av. de
  l'Observatoire, Paris, France}
  ,altaddress={Universit\'e de Picardie Jules Verne (UPJV), 33 Rue St Leu, Amiens, France} 
}

\author{L. Monaco}{
  address={European Southern Observatory, Casilla 19001, Santiago, Chile}
}

\author{S. Villanova}{
  address={Departamento de Astronomia, Casilla 160, Universidad de Concepcion, Chile}
}
\author{M. Catelan}{
  address={Departamento de Astronom\'ia y Astrof\'isica, Pontificia Universidad Cat\'olica de Chile, Av. Vicuna Mackena 4860, 782-0436 Macul, Santiago, Chile}
}
\author{P. Bonifacio}{
  address={2GEPI, Observatoire de Paris, CNRS, Univ. Paris Diderot, Place JulesJanssen, F92190 Meudon}
}
\author{M. Bellazzini}{
  address={INAF-Osservatorio Astronomico di Bologna, Via Ranzani 1, 40127, Bologna, Italy}
}
\author{C. Moni Bidin}{
  address={Departamento de Astronomia, Casilla 160, Universidad de Concepcion, Chile}
}
\author{G. Marconi}{
  address={European Southern Observatory, Casilla 19001, Santiago, Chile}
}
\author{D. Geisler}{
  address={Departamento de Astronomia, Casilla 160, Universidad de Concepcion, Chile}
}
\author{L. Sbordone}{
  address={Zentrum fur Astronomie der Universit\"at Heidelberg, Landessternwarte, Heidelberg, Germany}
  ,altaddress={GEPI, Observatoire de Paris, CNRS, Univ. Paris Diderot, Meudon, France}
}

\begin{abstract}
 We report preliminary results concerning the detailed chemical composition of metal poor stars belonging to close ultra-faint dwarf galaxies (hereafter UfDSphs). The abundances have been determined thanks to spectra obtained with  X-Shooter,  a high efficiency spectrograph installed on one of the  ESO VLT units.  
The sample of  ultra-faint dwarf spheroidal stars have abundance ratios slightly lower to what is measured in field halo star of the same metallicity.  We did not find extreme abundances in our Hercules stars as the one  found by Koch for his 2 Hercules stars.  The synthesis of the neutron capture elements Ba and Sr  seems to originate from the same nucleosynthetic process in operation during the early stages of the galactic evolution.

\end{abstract}

\maketitle


\section{Introduction}

While the discovery of the UfDSphs certainly alleviates the missing satellite problem, these loose objects are not expected to be important contributors to the mass assembly of the Milky Way (herafter MW) halo (\cite{Robertson2005},  \citep{Johnston2008}). On the other hand, being among the most metal-poor environments known ([Fe/H]< -2, \citep{Kirby2008}), UfDSphs may have been an important source for the building-up of the metal-poor end of the MW halo metallicity distribution (\citep{Frebel2010}, \citep{Kirby2008}, see Fig. 1).The chemical abundance patterns of UfDSph stars should then be similar to those of their halo counterparts at similar metallicity. This is a prediction and a crucial observational test for the hierarchical galaxy formation scenario. Detailed chemical patterns have been derived so far only for the UfDSphs Hercules (Her, two stars with [Fe/H]  close to -2.0, \cite{Koch2008}), Ursa Major II and Coma Berenices (UMa~II, Coma~Ber, 3 stars each,-3.2< [Fe/H]< -2.3, \cite{Frebel2010}).
These early works provide puzzling results. UfDSph stars have [alpha/Fe] abundance ratios either following the trend of MW stars (UMa II and ComBer) or that of brighter dwarf spheroidals (DSphs) of comparable metallicities. Neutron-capture elements (e.g. Ba and Sr) are, on the other hand, depleted in UfDSph stars compared to the common trend shared between MW and DSph stars. Therefore, it is still unclear whether UfDSphs can be flagged as building blocks of the metal-poor tail of the MW halo, and more observational data is mandatory to trace a self-consistent scenario for the formation, chemical enrichment history and evolution of both these fascinating systems and the MW itself.
 
\section{Observations}
 Available low-resolution studies provide membership and tentative metallicities for a number of stars in the UfDSphs. We made  use of the high efficiency, wide spectral coverage and relatively high resolution of X-shooter/VLT to perform a survey to derive detailed chemical abundances for a sample of 11 stars in 5 (out of the 9) UfDSphs accessible from Chile. All of the targets are UfDSph confirmed radial velocity members.X-Shooter is a multi-wavelength medium resolution spectrograph mounted at the UT2 Cassegrain focus. The instrument consists of three arms : UVB (300-560 nm), VIS (560-1024 nm) and NIR (1024-2480 nm). This paper presents premiminary results of a study of giants stars belonging to the UfDSphs Hercules, Canis Venatici II  and Leo IV.

We use the photometry information for BVI colors from the paper of Kirby et al. 2008 \cite{Kirby2008}. Temperature have been derived following the relations of Ramirez and Melendez (2005) \cite{Ramirez2005}. The surface gravities have been obtained from the photometry and calculated using the standard relation between logg, Mass, Teff and Mbol relative to the Sun, assuming the solar values  Teff=5790K, logg = 4.44 and Mbol= 4.72. We assumed also  $0.8 M_{\odot}$  for the mass of the giant stars which have been observed. Distance modulii have been taken in Belokurov et al. (2007)\cite{Belokurov2007}.

\section{Results and Conclusions}

\begin{figure}
\includegraphics[ height=.48 \textheight]{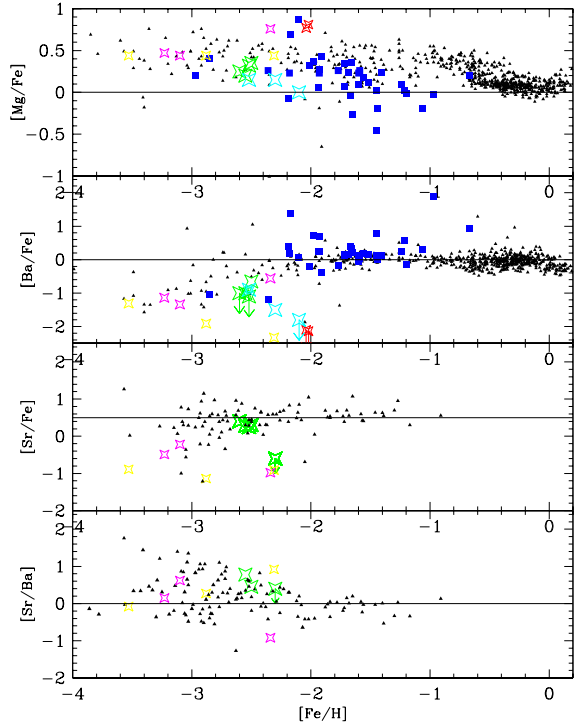}
\caption{ Black triangles: Halo field stars  \cite{Francois2007}, \cite{Venn2004}; blue rectangles: stars in DSphs \cite{Venn2004};
 purple star symbol: UMa~II  \cite{Frebel2010}; yellow star symbols : Coma~Ber \cite{Frebel2010}; 
 cyan star symbols: our data for Canis Venatici II and Leo IV UfDsph; red star symbols: stars in Herc UfDSph \cite{Koch2008}; green star symbols: our data for  Herc UfDsph.}
\end{figure}

Figure~1 shows the results obtained for our program stars for the ratio [Mg/Fe], [Sr/Fe] and [Ba/Fe] together with data collected in the literature for MW field stars \citep{Francois2007}, Dwarf Spheroidal (herafter DSph) stars \citep{Venn2004} and stars belonging to UfDSphs (\citep{Frebel2010},\citep{Koch2008}).  The ratio [Mg/Fe] is in  agreement with the values found in DSph stars. This value is slightly lower than the mean value found in the the metal poor end of the Milky Way. For [Ba/Fe], we found results similar, although slightly lower,  to what is found in field stars of our Galaxy. It should be noted that the larger difference is found for stars belonging to Coma~Ber. (Frebel et al. 2010,\cite{Frebel2010}). Our [Sr/Fe] results  are similar to the abundance ratios found in the field stars of the Milky Way. Again, the only objects which deviate significantly from the MW trend are the stars belonging to Coma~Ber. In the lowest panel of Fig.1  are plotted the ratios [Sr/Ba] as a function of [Fe/H].  This plot is particularly interesting as it reveals the variation of the ratio of an element with an early r-process nucleosynthesis (Ba) and an another early r-process element (Sr) built also with an additional r-process synthesis (weak r). This figure shows that the peculiar trend found  in the very metal poor stars  of our Galaxy is shared with the metal poor stars found  in the ultra faint dwarf galaxies. Only one star deviates significantly from the trend.   

Although the full  sample has not been analyzed completely, it appears that UfDSph stars have abundance ratios slightly lower to what is measured in field halo star of the same metallicity.  We did not find extreme abundances in our Hercules stars as the one  found by Koch \citep{Koch2008}.  The synthesis of the neutron capture elements Ba and Sr  in the UfDSphs seems to originate from the same nucleosynthetic process in operation during the early stages of the galactic evolution.





\bibliographystyle{aipproc}   


\end{document}